\documentclass[aps,prb,twocolumn,floatfix,amsmath,amssymb]{revtex4}
\usepackage[final]{graphicx}
\usepackage{bm}
\usepackage{afterpage}
\usepackage{color}
\usepackage{comment}
\usepackage[colorlinks=true,urlcolor=blue,linkcolor=blue,citecolor=blue]{hyperref}

\emergencystretch=\maxdimen
\begin{document}

\title{Quantum Monte Carlo study of honeycomb antiferromagnets under a triaxial strain}

\author{Junsong Sun}
\affiliation{School of Physics, Beihang University,
Beijing, 100191, China}

\author{Nvsen Ma}
\email{nvsenma@buaa.edu.cn}
\affiliation{School of Physics, Beihang University,
Beijing, 100191, China}

\author{Tao Ying}
\affiliation{Department of Physics, Harbin Institute of Technology, 150001 Harbin, China}

\author{Huaiming Guo}
\email{hmguo@buaa.edu.cn}
\affiliation{ School of Physics, Beihang University,
Beijing, 100191, China}
\affiliation{Beijing Computational Science Research Center, Beijing 100193, China}

\author{Shiping Feng}
\affiliation{ Department of Physics,  Beijing Normal University, Beijing, 100875, China}

\begin{abstract}
The honeycomb antiferromagnet under a triaxial strain is studied using the quantum Monte Carlo simulation. The strain dimerizes the exchange couplings near the corners, thus destructs the antiferromagnetic order therein. The antiferromagnetic region is continuously reduced by the strain. For the same strain strength, the exact numerical results give a much smaller antiferromagnetic region than the linear spin-wave theory. We then study the strained $XY$ antiferromagnet, where the magnon pseudo-magnetic field behaves quite differently. The $0$th Landau level appears in the middle of the spectrum, and the quantized energies above (below) it are proportional to $n^{\frac{1}{3}} (n^{\frac{2}{3}})$, which is in great contrast to the equally-spaced ones in the Heisenberg case. Besides, we find the antiferromagnetic order of the $XY$ model is much more robust to the dimerization than the Heisenberg one. The local susceptibility of the Heisenberg case is extracted by the numerical analytical continuation, and no sign of the pseudo-Landau levels is resolved. It is still not sure whether the result is due to the intrinsic problem of the numerical analytical continuation. Thus the existence of the magnon pseudo-Landau levels in the spin-$\frac{1}{2}$ strained Heisenberg Hamiltonian remains an open question. Our results are closely related to the two-dimensional van der Waals quantum antiferromagnets, and may be realized experimentally.
\end{abstract}

\pacs{
  71.10.Fd, 
  03.65.Vf, 
  71.10.-w, 
}

\maketitle

\section{Introduction}

Mechanical deformation is a powerful approach to tailor the electronic properties of quantum materials. The most dramatic examples include: applying various kinds of strains to monolayer graphene and twisting multilayer graphene. Remarkably, the recent discovery of the unconventional superconductivity in twisted bilayer graphene has made a perfect playground for exploring correlated physics\cite{Bistritzer2011,cao2018,cao2018b}, and even creates a new research field called 'twistronics'\cite{kaxiras2017}.

Strain engineering in the context of graphene has a longer history, and has been extensively investigated both theoretically and experimentally.
Strain-induced pseudo-magnetic field was first discovered in graphene nanobubbles grown on the platinum surface\cite{Levy544}. The observed Landau levels correspond to a field strength greater than 300 Tesla, which is far beyond the current lab condition and provides a unique opportunity for studying the physics in extremely high magnetic field regime.
The high pseudo-magnetic fields have also been achieved in subsequent experiments by placing graphene on curved or predesigned substrates\cite{Klimov2012,helin2015,Niggeeaaw2019,Hsu2020,RN45}.
However the above pseudo-magnetic fields are in very localized and curved regions. It is highly desirable to realize a uniform pseudo-magnetic field in a planar sheet for the purpose of mimicking the physical phenomena under real magnetic fields. Consequently, various methods of engineering the corresponding strains have been proposed, such as: triaxial strain\cite{guinea2010,peeters2013,mikkel2016}, bending graphene ribbons\cite{novoselov2010,tony2010,Inhomogeneous2017}, uniaxial strain increasing linearly in the applied direction\cite{helin2013,castro2017,franz2020}, uniaxial strain on a shaped graphene ribbon\cite{liteng2015}, et.al..

Since the pseudo-magnetic field is generated by simply modulating the strengths of the hopping amplitudes using strain, it can be directly extended to charge-neutral particles, which opens a door to explore the novel physical properties under the magnetic field that can not be studied elsewhere.
Indeed strain-induced Landau levels have been recently generalized to neutral Bogoliubov particles in nodal superconductors\cite{arun2017,nica2018}, Majorana fermions in spin liquids\cite{rachel2016}, and magnons in quantum magnets\cite{yago2018,vojta2019,liu2020}.
Although similar pseudo-Landau levels (PPLs) are formed, they are quite different from their counterparts in graphene. For example, the magnon PLLs are on top of the magnon spectrum for the strained honeycomb antiferromagnets. Besides, they are equally spaced, which is in contrast to the $\sqrt{n}$ behavior of Dirac fermions\cite{vojta2019,RevModPhys.83.1193}.

Although an abundance of interesting physics are expected therein, it is still challenging to deal with the underlying strongly-correlated systems. It is noted that the existing studies only treat the strained quantum spin systems within the linear spin-wave theory (LSWT). Since quantum fluctuations scale as $1/S$ for systems with bilinear spin interactions, LSWT becomes accurate only for large values of the spin $S$. Hence for spin-$\frac{1}{2}$ strained Heisenberg Hamiltonians, it is natural to ask to what extend LSWT makes the approximation and whether the spin-wave results persist in a rigorous calculation.

In this manuscript, we study the Heisenberg and $XY$ Hamiltonians on the honeycomb lattice under a triaxial strain utilizing LSWT and QMC method. First we address the evolution of the AF order, which is characterized by a finite local magnetization and long-range AF correlations. Both approaches show the AF order begins to vanish from the corners, and persists in the central region of the triangular geometry. The size of the AF region is continuously reduced with increasing the strain. The realistic size obtained by QMC is much smaller than that of LSWT. We also utilize QMC to calculate the imaginary time correlations of the spins, and extract the local susceptibility with the numerical analytical continuation. We find no sign of PLLs in the obtained $S(i,i,\omega)$. This poses the suspicion of the existence of PLLs beyond LSWT. In the last part of the manuscript, we present the results of the strained $XY$ Hamiltonian. While the evolution of quantum antiferromagnetism with strain strength is similar to the Heisenberg case, the properties of the magnon PLLs are quite different. These results are closely related to the two-dimensional van der Waals quantum antiferromagnets, and will attract both theoretical and experimental interests.

This paper is organized as follows. Section II introduces the precise
model we will investigate, along with our computational methodology.
Section III presents the LSWT results of the strained Heisenberg model. Section IV uses QMC simulations to study the evolution of the quantum antiferromagnetism and resolve PLLs. Section V demonstrates the LSWT and QMC results of the $XY$ Hamiltonian under a triaxial strain. Section VI contains some further discussions and the conclusions.

\section{The model and method}

\begin{figure}[htbp]
\centering \includegraphics[width=9.cm]{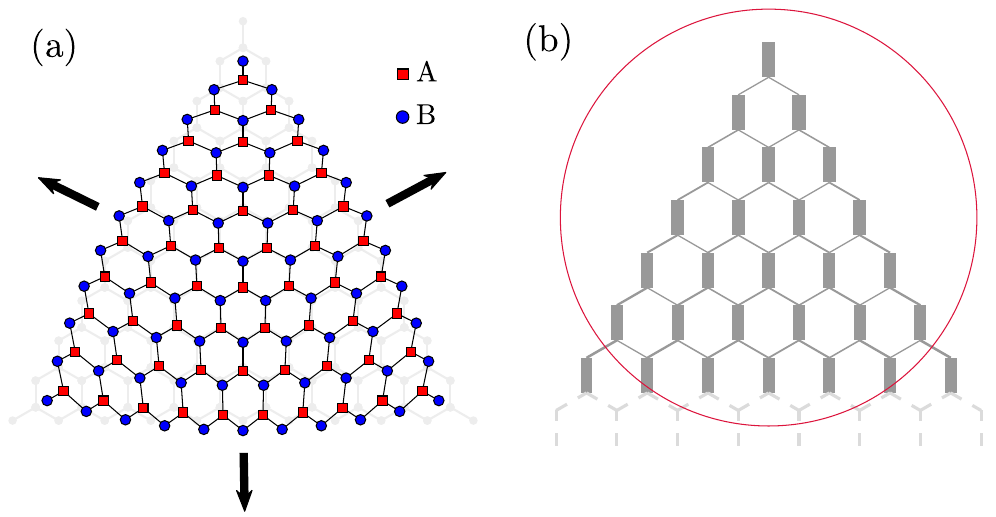} \caption{(a) Deformed honeycomb lattice with a triangular shape under a triaxial strain. The lattice has $L^2$ sites with the linear size $L=11$ for the figure. The number of sites in sublattices A and B are unequal with $N_A=L(L-1)/2$ and $N_B=L(L+1)/2$. (b) Enlarged plot in the vicinity of the upper corner of a $L=50$ triangular geometry. The value of the exchange coupling on each bond is represented by the thickness of the bond. $c/c_{max}=0.8$ is used for the strain strength in (b).}
\label{fig1}
\end{figure}

We consider an antiferromagnetic Heisenberg model on the honeycomb lattice, which in the absence of strain writes as
\begin{align}
H_0=J\sum_{\langle ij\rangle}{\bf S}_i\cdot{\bf S}_j,
\end{align}
where $J$ is the antiferromagnetic exchange coupling; ${\bf S}_{i}=(S_i^{x},S_i^{y},S_i^{z})$ is spin-$\frac{1}{2}$ operator on the site $i$, which obeys commutation relations, $\left[S^{a_1}_{i},S^{a_2}_{j}\right]=i\hbar \varepsilon _{a_1a_2a_3}S^{
a_3}_{i}\delta_{ij}\delta_{{\bf r},{\bf r'}}$ with $\varepsilon _{a_1a_2a_3}$ the Levi-Civita symbol and $a_1,a_2,a_3= x,y,z$ representing spin components.

In the context of graphene under strain, the empirical relation of the hopping amplitude is $t_{ij}=te^{-\beta \tau_{ij}}$, where $\beta$ is the Gr\"{u}nisen parameter and $\tau_{ij}$ is the relative change in the length of the bond connecting the sites $i,j$. The AF Heisenberg model can be derived based on the one-band Hubbard model in the large-$U$ limit, where the exchange coupling is $J\propto t^2/U$ with $U$ the on-site Hubbard interaction. Thus in the presence of strain, the exchange coupling also varies in an exponential manner, i.e., $J_{ij}=Je^{-2\beta \tau_{ij}}$.
Here we adopt a linear approximation to the exponential function, and the exchange couplings in the Hamiltonian becomes,
\begin{align}
J\longrightarrow J_{ij}=J[1-\gamma(\frac{|\vec{\delta}_{ij} |}{a_0}-1)],
\end{align}
where $a_0$ is the lattice constant of the undeformed honeycomb lattice; $\gamma$ is the Gr\"{u}neisen parameter and represents the strength of magnetoelastic coupling ($\gamma=1000$ is used throughout the paper); $\vec{\delta}_{ij}$ is the vector connecting the $i$-th and $j$-th sites after the deformation, which can be calculated with the displacement $\vec{u}({\bf r})=[u_x({\bf r}),u_y({\bf r})]$ of the lattice site at position ${\bf r}=(x,y)$,
\begin{align}
\vec{\delta}_{i j}=\vec{\bf r}_{i}+\vec{u}\left({\bf r}_i\right)-[\vec{\bf r}_{j}+\vec{u}\left({\bf r}_j\right)].
\end{align}
For large lattice, the displacement can be taken as a smooth function of the coordinates, and the strain tensor is defined as
$$
\epsilon_{i j}=\frac{1}{2}\left[\partial_{j} u_{i}+\partial_{i} u_{j}\right], \quad i, j=x, y.
$$
The three nearest-neighbor bond lengths becomes $a_0(1+\Delta u_{n})(n=1,2,3)$ with
\begin{align}
\Delta u_{n}=\sum_{i, j} \frac{a_{n}^{i} a_{n}^{j}}{a_{0}^{2}} \epsilon_{i j},
\end{align}
and $\vec{a}_n$ the vector of the original bond.
In graphene, such a spatial modulation leads to a gauge field
\begin{align}
\mathbf{A}=\frac{\gamma}{2}\left(\begin{array}{c}
\epsilon_{x x}-\epsilon_{y y} \\
-2 \epsilon_{x y}
\end{array}\right).
\end{align}
To generate an almost homogeneous pseudo-magnetic field, one of the commonly adopted approaches is the triaxial strain applied to a honeycomb flake, whose displacement field is given by\cite{guinea2010}
\begin{align}
\vec{u}({\bf r})=c(2xy,x^2-y^2),
\end{align}
where $c$ measures the strain strength. In the presence of strain, the vertical bonds on the outmost sites of the zigzag edges have the smallest exchange couplings, whose values decrease with increasing the strain strength. Since the coupling value should be positive, the zero value determines a maximum strain that can be applied to the triangular geometry with a fixed size. Similarly for a fixed strain strength, when the lattice is large enough, the bonds mentioned above vanish and a zigzag-terminated triangular geometry is cut out from the whole lattice, naturally setting a limiting size for the triangular geometry\cite{PhysRevB.90.155418}.

In the following discussions, we study the model in Eq.(1) under the above triaxial strain. We first use LSWT to get insights, and then employ the approach of stochastic series expansion (SSE) QMC method~\cite{sandvik2002,syljuasen2003} with directed loop updates to perform exact numerical simulations. The SSE method expands the partition function in power series and the trace is written as a sum of diagonal matrix elements. The directed loop updates make the simulation very efficient~\cite{Bauer2011,fabien2005,pollet2004}. Our simulations are on a triangular geometry with the total number of sites $N_s=L(L+1)$ with $L$ the linear size. There are no approximations
causing systematic errors, and the discrete configuration space can be sampled without floating
point operations. The temperature is set to be $\beta=50$, which is low enough to obtain the ground-state properties. For such spin systems on bipartite lattice, the notorious sign problem in the QMC approach can be avoided.

\section{The linear spin wave theory}
Let us first get some insights into the properties of the model Eq.(1) by the LSWT analysis.
The Heisenberg model under a triaxial strain can be treated within LSWT by replacing the spin operators by bosonic ones via Holstein-Primakoff (HP) transformation\cite{hptransformation}. The transformation on sublattice A (the spin is in the positive $z$-direction) is defined as
\begin{align}\label{eq2}
S^+_{i}&=\sqrt{2S}a_{i}, S^-_{i}=\sqrt{2S}a^{\dagger}_{i},\\ \nonumber
S^z_{i}&=S-a^{\dagger}_{i}a_{i}.
\end{align}
On sublattice B (the spin is in the negative $z$-direction), the spin operators are defined as
\begin{align}\label{eq3}
S^+_{i}&=\sqrt{2S}b^{\dagger}_{i},S^-_{i}= \sqrt{2S}b_{i},\\ \nonumber
S^z_{i}&=b^{\dagger}_{i} b_{i}-S.
\end{align}
Keeping only the bilinear terms, the bosonic tight binding Hamiltonian reads
\begin{align}
H=\sum_{\langle i j\rangle} J_{ij} S\left(a_{i} b_{j}+a_{i}^{\dagger} b_{j}^{\dagger}+a_{i}^{\dagger} a_{i}+b_{j}^{\dagger} b_{j}\right).
\end{align}
Under the basis $X^{\dagger}=(a_1^\dagger ,...,a_{N_A}^\dagger ,b_1,...,b_{N_B})$, the above Hamiltonian writes compactly as $H=X^{\dagger}MX$, where $M$ is a $N_s\times N_s$ matrix with $N_s=N_A+N_B$ the total number of sites.
By a Bogliubov transformation\cite{PhysRev.139.A450,xiao2009theory}, the matrix $M$ becomes diagonal,
\begin{align}
T^\dagger MT=\rm{diag}(\omega_1,...,\omega_{N_s}),
\end{align}
where $T$ has the following form
\begin{align}\label{}
  T=\left(
      \begin{array}{cc}
        A_+ & A_- \\
        B_+ & B_- \\
      \end{array}
    \right),
\end{align}
with $A_{+}$ a $N_A\times N_A$ matrix and $B_{-}$ a $N_B\times N_B$ one.
Then the diagonal basis is connected to the original one by
\begin{align}
\left[\begin{array}{c}
a_{1} \\
\vdots \\
a_{N_{A}}
\end{array}\right]&=A_{+}\left[\begin{array}{c}
\alpha_{1} \\
\vdots \\
\alpha_{N_{A}}
\end{array}\right]+A_{-}\left[\begin{array}{c}
\beta_{1}^{\dagger} \\
\vdots \\
\beta_{N_{B}}^{\dagger}
\end{array}\right], \\ \nonumber
\left[\begin{array}{c}
b_{1}^{\dagger} \\
\vdots \\
b_{N_{B}}^{\dagger}
\end{array}\right]&=B_{+}\left[\begin{array}{c}
\alpha_{1} \\
\vdots \\
\alpha_{N_{A}}
\end{array}\right]+B_{-}\left[\begin{array}{c}
\beta^{\dagger}_{1} \\
\vdots \\
\beta^{\dagger}_{N_{B}}
\end{array}\right].
\end{align}

Within LSWT, several physical quantities can be deduced.
The local magnetization can be directly calculated as,
\begin{align}
\left|\left\langle S_{i \in A}^{z}\right\rangle\right|&=S-\sum_{k=1}^{N_A}\left(A^{*}_{-,i k} A_{-,i k}\right), \\ \nonumber
\left|\left\langle S_{i \in B}^{z}\right\rangle\right|&=S-\sum_{k=1}^{N_B}\left(B^{*}_{+,i k} B_{+,i k}\right).
\end{align}

The appearance of sharp peaks in the magnon density of states is direct evidence of the formation of PLLs. With the magnon spectrum obtained by LSWT, the magnon density of states can be calculated as
\begin{align}
 \rho(\omega)=\sum_k\delta(\omega-\omega_k)=\sum_k \frac{1}{\sqrt{2\pi}c_0}e^{-\frac{(\omega-\omega_k)^2}{2c_0^2}},
\end{align}
where $\delta$ function is approximated by a narrow Gaussian wave packet with $c_0$ a small constant.

In the many-body calculations, the local susceptibility is a more relevant quantity, which is defined as\cite{PhysRevB.71.104427}
\begin{align}
S(i, i, \omega)=\int_{-\infty}^{\infty} d t e^{i \omega t}\left\langle S_{i}^{x}(t) S_{i}^{x}+S_{i}^{y}(t) S_{i}^{y}\right\rangle.
\end{align}
Within LSWT, the following expression for $\chi_{loc}=S(i, i, \omega)$ is obtained,
\begin{align}
\chi^{A}_{loc}=&\sum_{k=1}^{N_B}A_{-,ik} A_{-,i k}^{*}\delta\left(\omega-\omega_{k}\right)\\ \nonumber
              +&\sum_{k=1}^{N_A}A_{+,i k}^{*} A_{+,i k} \delta\left(\omega-\omega_{k}\right),
\end{align}
when $i$ is in sublattice A, and for $i$ in sublattice B
\begin{align}
\chi^{B}_{loc}=&\sum_{k=1}^{N_B}B_{+,i k} B_{+,i k}^{*}\delta\left(\omega-\omega_{k}\right)\\ \nonumber
              +&\sum_{k=1}^{N_A}B_{-,i k}^{*} B_{-,i k} \delta\left(\omega-\omega_{k}\right).
\end{align}
The above formulas show that the local susceptibility is equivalent to the density of states in characterizing the peaks corresponding to the flat PLLs. More importantly, the local susceptibility can be exactly determined by numerical analytic continuation of the imaginary time spin correlations obtained by the QMC simulation.

We first study how the AF order is affected by the strain. Figure 2 shows the distribution of the local magnetization at several values of the strain strength. In the absence of strain, although the N{\'e}el order is perturbed by the open boundaries, the long-range AF order preserves, and the profile of $m_s(i)$ remains almost uniform. We calculate the spatial averaged magnetization,
\begin{align}
m_s=\frac{1}{N_s}\sum_{1}^{N_s}m_s(i).
\end{align}
The value is $m_s=0.218$, which is about $80$ percent of the QMC value ($0.271$). After the strain is applied, the sites near the boundaries are altered. Especially the local magnetizations near the three corners change dramatically, where the value of $m_s(i)$ even becomes negative, implying the AF order vanishes there. In contrast, the central region of the triangle geometry is less affected, where the distribution remains uniform, and the magnitudes do not change much. As the strain strength increases, the size of the central AF region shrinks, and the AF order is destructed on more sites near the boundaries.
It should be noted that the appearance of negative $m_s(i)$ is meaningless in LSWT since LSWT is a perturbation theory based on the AF order. Nevertheless, it can still qualitatively reflect the evolution of the AF order under an applied strain. In the following section, the exact results will be obtained using the QMC simulations.

\begin{figure}[htbp]
\centering \includegraphics[width=9.cm]{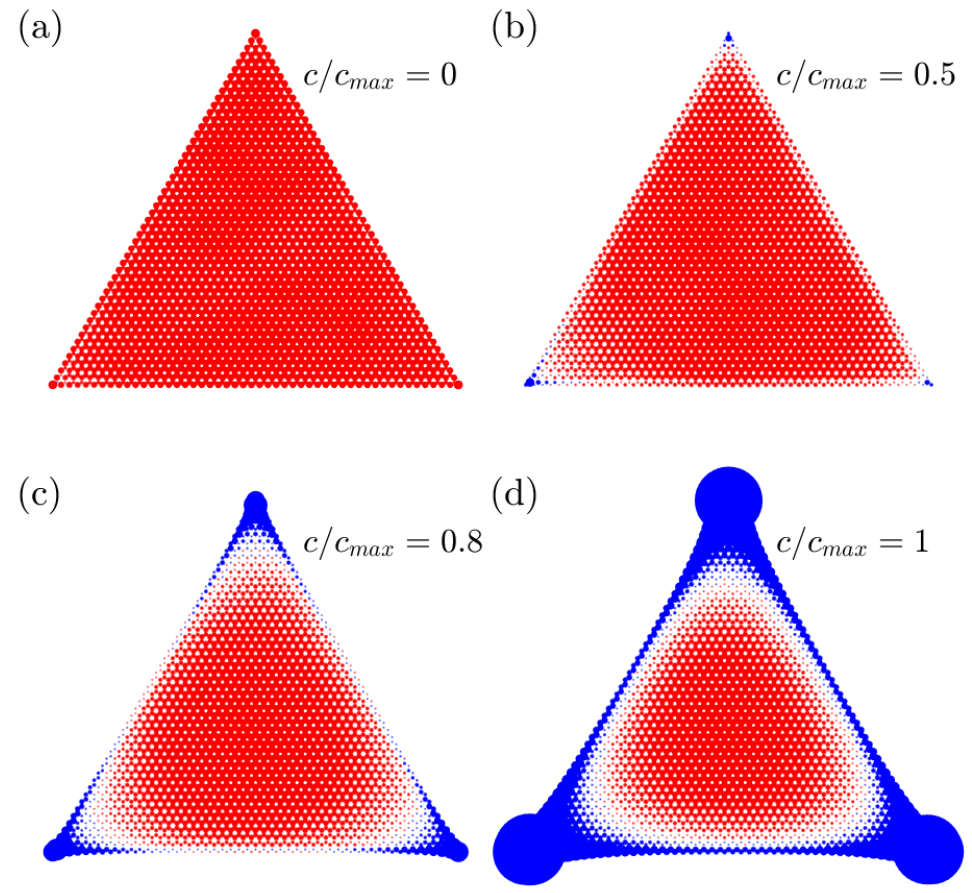} \caption{The distribution of the local magnetization obtained with the LSWT at several values of the strain strength $c/c_{max}=0,0.5,0.8,1$. Red (blue) color represents positive (negative) values, and the values are represented by the radii of the circles. Here the linear size of the lattice is $L=50$.}
\label{fig2}
\end{figure}

The magnon spectrum for the open triangular geometry is shown in Fig.3(a), which consists of discrete degenerate sets of energy levels. It has been deduced in Ref. 25 that the energy levels can be approximated by $\omega_n=3J(1-n/L)$ with degeneracy $d_0=L-1,d_L=1$, and $d_n=2(L-n)$ for $n=1,...,L-1$. The approximation is found to agree well with the tight-binding calculations for large $L$ and small $n$. The spectrum spans a range up to $3JS$, which does not change when altering the strain. Near the upper end of the spectrum, a series of discrete plateaus are clearly resolved. Figure 3(b) plots the differences of the adjacent energy eigenvalues, where the plateau width is demonstrated by the distance between two adjacent peaks. It shows that the plateaus have approximately the same widths except the highest one, whose width is about half of the other ones. As the energy levels are far away from the upper end, although the plateaus are tending to become less apparent and the peaks in Fig.3(b) are broadened, the distances between adjacent peaks change little.
Besides, the gaps between adjacent energy levels are almost the same, which is best manifested in the magnon density of states, where equally-spaced peaks are clearly visible [see Fig.3(c)]. These observations imply the applied strain induces a pseudo-magnetic field, resulting in the equally-spaced magnon PLLs starting from the upper end of the spectrum. These equally-spaced PLLs are also reflected in the periodic oscillating behavior of the local susceptibility, as shown in Fig.3(e) and (f). Here it is noted that the $n=0$ PLL is only located on sublattice A, which is similar to the case of strained graphene.

For honeycomb antiferromagnets without strain, the low-energy spin-wave excitations display a linear spectrum, which results in a linear magnon density of states. This behavior can be best seen from the integrated density of states,
\begin{align}
N(\omega)=\int_{0}^{\omega}\rho(\epsilon)d\epsilon.
\end{align}
This quantity has a more smooth curve than $\rho(\omega)$ due to the integration. As shown in Fig.3(d), $N(\omega)$ shows an expected quadratic low-energy behavior corresponding to the linear magnon density of states. In the presence of strain, $N(\omega)$ remains linear over $\omega^2$, but the slope increases as the strain is strengthened. Since $\rho(\omega)$ is proportional to the inverse of the slope of the magnon dispersion,  the low-energy excitation spectrum is gradually flattened by the strain.

\begin{figure}[htbp]
\centering \includegraphics[width=8.8cm]{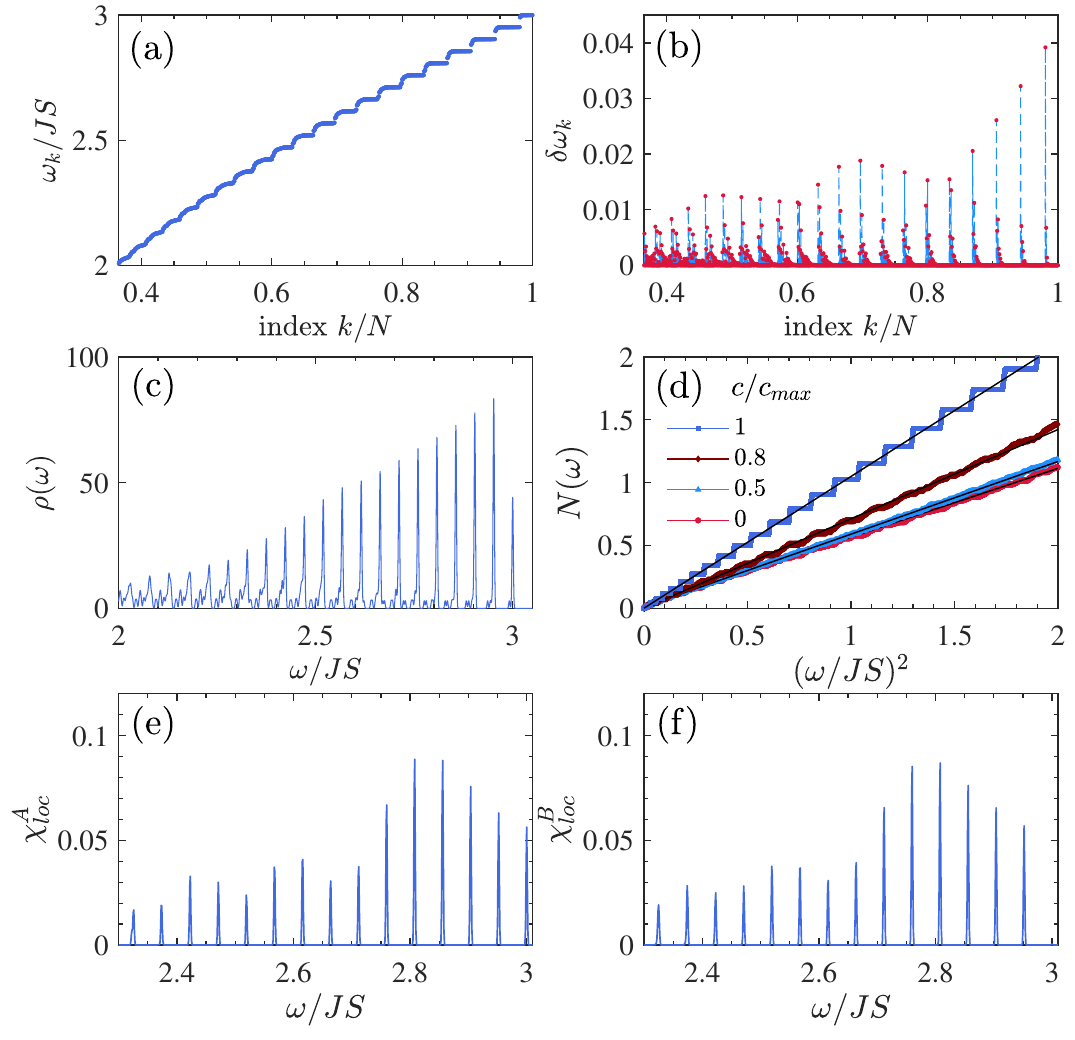} \caption{(a) The magnon spectrum near the upper end of the energy as a function of the normalized index. (b) The corresponding differential of the energy levels. (c) The magnon density of states for the energy spectrum in (a). (d) Integrated density of states $N(\omega)$ for several values of the strain strength. Local susceptibility at the sample center on: (e) A sublattice and (f) B sublattice. In (e) and (f), only the results at the upper ends of the energy are shown, where PLLs are most apparent. The linear size is $N=50$. In (a-c,e,f) the strain strength is $c/c_{max}=0.8$.}
\label{fig3}
\end{figure}

\section{The QMC results}

We first calculate the internal energy per site. As shown in Fig.4(a), the average energy decreases as the strain strength is increased. The applied strain generally makes the exchange couplings nonuniform. Figure 4(b) shows the distribution of $J_{ij}$ at a typical strength $c/c_{max}=0.8$, which extends between about $0.2$ and $2.6$ with a decreasing probability. The resulting average value of $J_{ij}$ increases monotonically with the strain strength. Since $\langle J_{ij}\rangle$ is regarded as the energy scale of the internal energy, the decrease of the internal energy can be qualitatively accounted for by the increase of $\langle J_{ij}\rangle$.

\begin{figure}[htbp]
\centering \includegraphics[width=6.5cm]{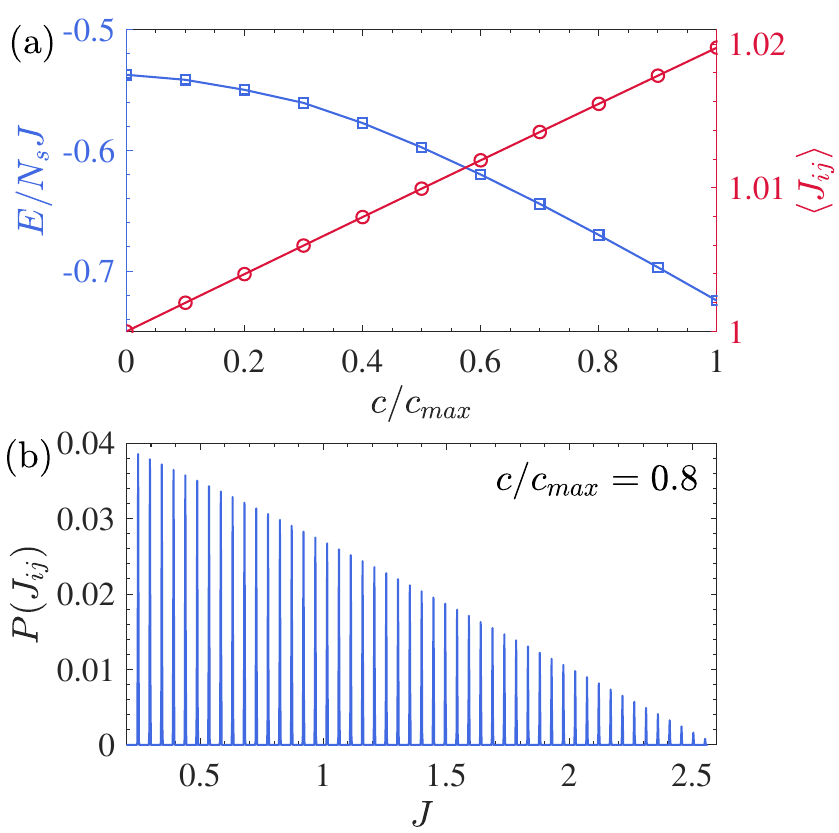} \caption{(a) The internal energy per site and the average value of the exchange couplings as a function of strain strength $c/c_{max}$. (b) The distribution probability of $J_{ij}$ at the strain strength $c/c_{max}=0.8$. The linear size is $L=50$. }
\label{fig4}
\end{figure}

In QMC simulations, the local value of the magnetization is proportional to $m_s^{qmc}(i)$, defined by\cite{PhysRevLett.90.177205}
\begin{align}
m_{s}^{qmc}(i)=\sqrt{\frac{3}{N} \sum_{j=1}^{N}\textrm{sgn}(i,j)\left\langle S_{i}^{z} S_{j}^{z}\right\rangle},
\end{align}
where the sum is over all lattice sites $j$, and $\textrm{sgn}(i,j)=1(-1)$ if $i,j$ belong to the same (opposite) sublattice. Figure 5 plots the distribution of $m_s^{qmc}(i)$ at the  same strain strengths as those in Fig.2. After the strain is applied, the values of $m_s^{qmc}(i)$ becomes negligible small near the three corners, and remain finite in the central region, implying the AF order preserves here. The AF region becomes reduced as the strain strength is enlarged. Although the evolution of the AF order with strain is consistent with the result from LSWT, the actual size of the AF region is much smaller than that the spin-wave one. This implies the LSWT overestimates the magnetic order.

\begin{figure}[htbp]
\centering \includegraphics[width=9.cm]{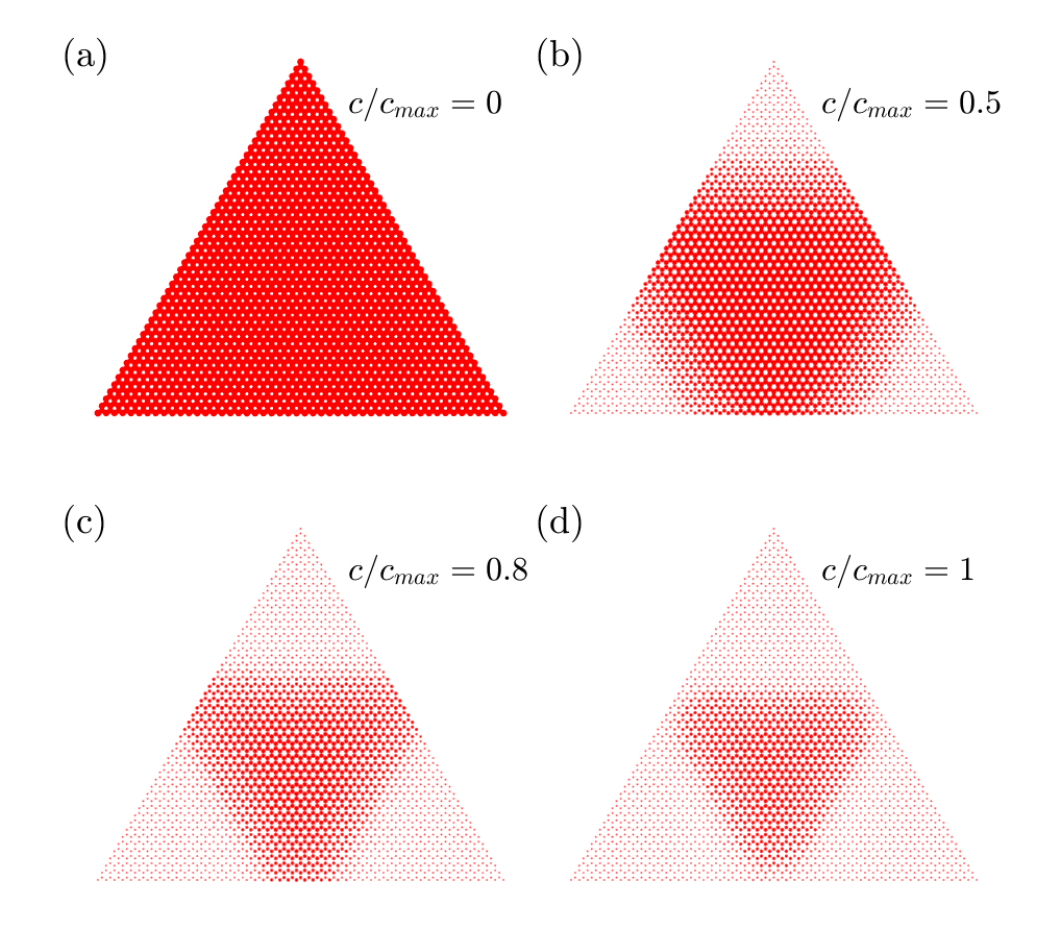} \caption{The distribution of $m_s^{qmc}(i)$ obtained with QMC simulations at the strain strength $c/c_{max}=0,0.5,0.8,1$. The values of $m_s^{qmc}(i)$ are represented by the radii of the red solid circles. The linear size is $L=50$.}
\label{fig5}
\end{figure}

We also calculate the static structure factor to the study the magnetic order, which is defined by,
\begin{align}
S({\bf k})=\frac{1}{N_s}\sum_{i,j=1}^{N_s}\langle S_i^zS_j^z\rangle e^{i{\bf k}\cdot ({\bf r}_i-{\bf r}_j)}.
\end{align}
Figure 6 plots the static structure factor at two typical strain strength $c/c_{max}=0,0.8$. Without the strain, the AF long-range order corresponds to sharp peaks appearing at ${\bf k}=m{\bf b}_1+n{\bf b}_2$, where ${\bf b}_1, {\bf b}_2$  are the reciprocal lattice vectors, and $m,n$ are arbitrary integers. After the strain is applied, the peaks become broader, but their positions are unchanged. This demonstrates that while the AF order is destructed by the strain on spatial average, no other new magnetic structures are generated.
The average magnetization is directly calculated as
\begin{align}
m_{s}^{qmc}=\sqrt{\frac{1}{N_s}\sum_{i}[m^{qmc}_{s}(i)]^2},
\end{align}
which amounts to the spin-wave average magnetization $m_{s}$. We find the QMC values quantitatively agree with the spin-wave ones, both of which decrease monotonically with increasing the strain.

\begin{figure}[htbp]
\centering \includegraphics[width=9.cm]{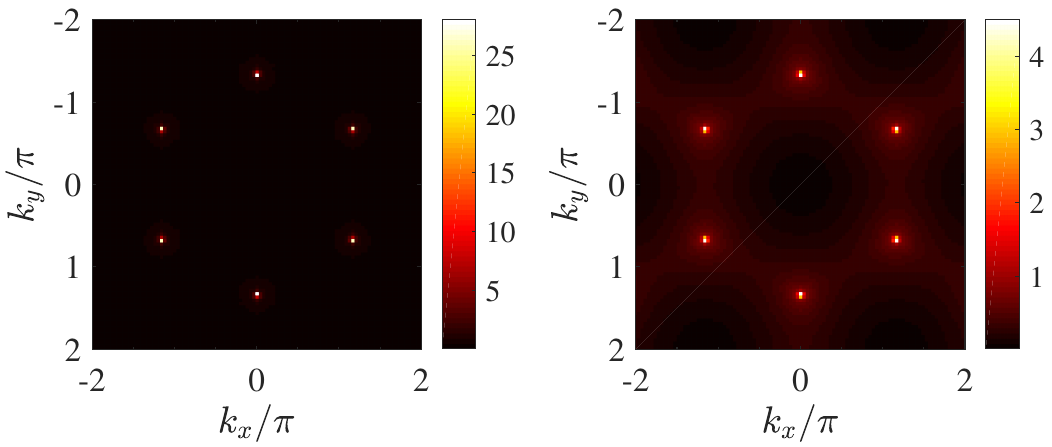} \caption{The false color plot of the static structure factor $S({\bf k})$ obtained with QMC at the strain strength $c/c_{max}=0$(left) and $0.8$(right). The linear size is $L=50$.}
\label{fig6}
\end{figure}

Next we examine the nature of the nonmagnetic phase induced by the strain near the three corners. As shown in Fig.1(b), the exchange couplings near the corners are periodically modulated, forming dimerizations along the two primitive directions of each corner. Hence the nonmagnetic phase is induced by the dimerizations, which should be gapped in the spin excitation. To see this explicitly, we add a perpendicular magnetic field described by
\begin{align}
H_{h}=-\sum_{i}hS_i^z.
\end{align}
The total Hamiltonian $H_0+H_{h}$ is then simulated by the QMC method, and we measure the magnetization of specific regions defined as $M_z^r=\sum_i^{N_r}S_i^z$ with $N_r$ the number of sites in a selected region. For a corner region, we find $M_z^{corner}=0$ below a critical magnetic field $h_c$, and above $h_c$, $M_z^{corner}$ begins to increase continuously. This behavior verifies the existence of a spin gap, and is consistent with the property of a dimerized phase. Figure 7 also plots the magnetization $M_z$ of the whole lattice. Interestingly, $M_z$ is quantized up to a finite magnetic field. The quantized value, $M_z=L/2$, is exactly half the difference of the number of sites in the two sublattices. This implies that the ground state has a total free spin,
\begin{align}
S_{tot}=\frac{1}{2}|N_A-N_B|,
\end{align}
which is induced by the imbalance of the two sublattices. The existence of a residual spin is originally established for the repulsive Hubbard model on imbalanced bipartite lattices\cite{Lieb1989,Yazyev_2010}. Here we show the theorem remains correct for the AF Heisenberg model. We also calculate the magnetization density $M_z^{central}$ in the central region, and find the curve differs little from that of the whole region. It demonstrates that the above free spin is mainly located in the central AF region.

\begin{figure}[htbp]
\centering \includegraphics[width=7.cm]{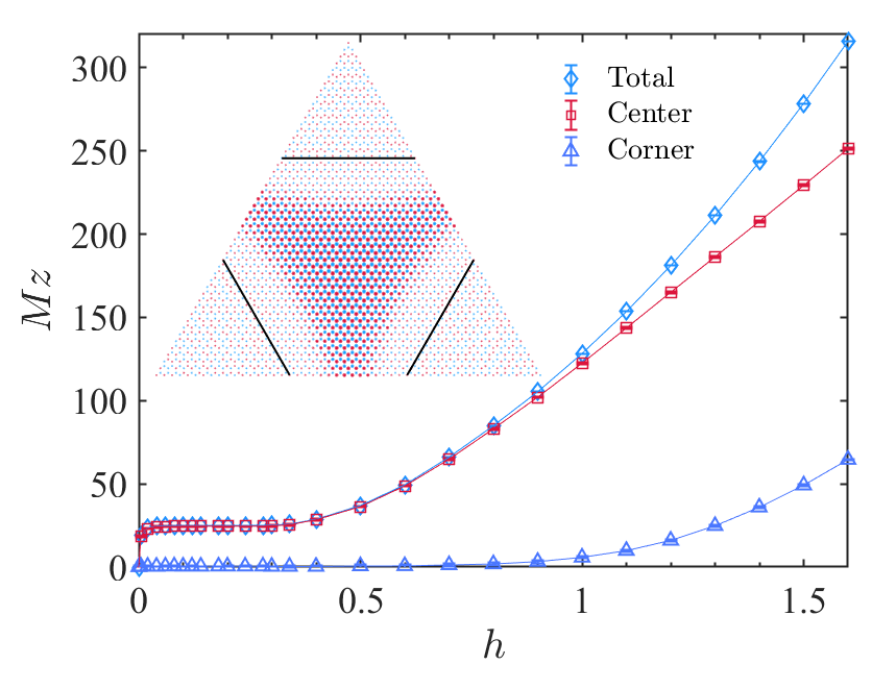} \caption{The magnetizations of different regions as a function of $h$. Inset shows the local distribution of $M_z$
induced by a magnetic field $h = 0.3$. The solid black lines in the inset mark the boundaries between the central and corner regions.}
\label{fig7}
\end{figure}

The QMC method can directly measure the imaginary time correlation of the $z$-component of the spins\cite{mahan2013many},
\begin{align}
S(i,i,\tau)=\langle S_i^z(\tau)S_i^z\rangle,
\end{align}
where $S_i^z(\tau)=e^{\tau H}S_i^ze^{-\tau H}$, and the imaginary time is in the range $[0,\beta)$. The local susceptibility in the real-frequency space is thus obtained by inverting the relationship,
\begin{align}
S(i,i,\tau)=\int d\omega e^{-\tau \omega}S(i,i,\omega).
\end{align}
Generally this inverse problem is ill-posed, and there is no closed-form solution for $S(i,i,\omega)$. We carry out the extraction using the stochastic analytical continuation (SAC)\cite{mark1991,sandvik2016,sandvik2017,nvsenma2018}. Figure 8 plot $\chi_{loc}^{A}$ and $\chi_{loc}^{B}$ at the sample center for the strain strength $c/c_{max}=0.5,0.8$. The extracted curves are continuous, and show no sign of PLLs.
However, it does not always work for any analytical continuation methods to get a spectrum function consists of many peaks or delta functions. We can not easily rule out the possiblity of the existence of PLLs from the SAC results so far.  More knowledge and better understanding of the spin-1/2 strained Heisenberg model are necessary by improving the SAC calculation. Thus, whether there are PLLs in spin-1/2 strained Heisenberg  model or not still remains an open question.

\begin{figure}[htbp]
\centering \includegraphics[width=8.cm]{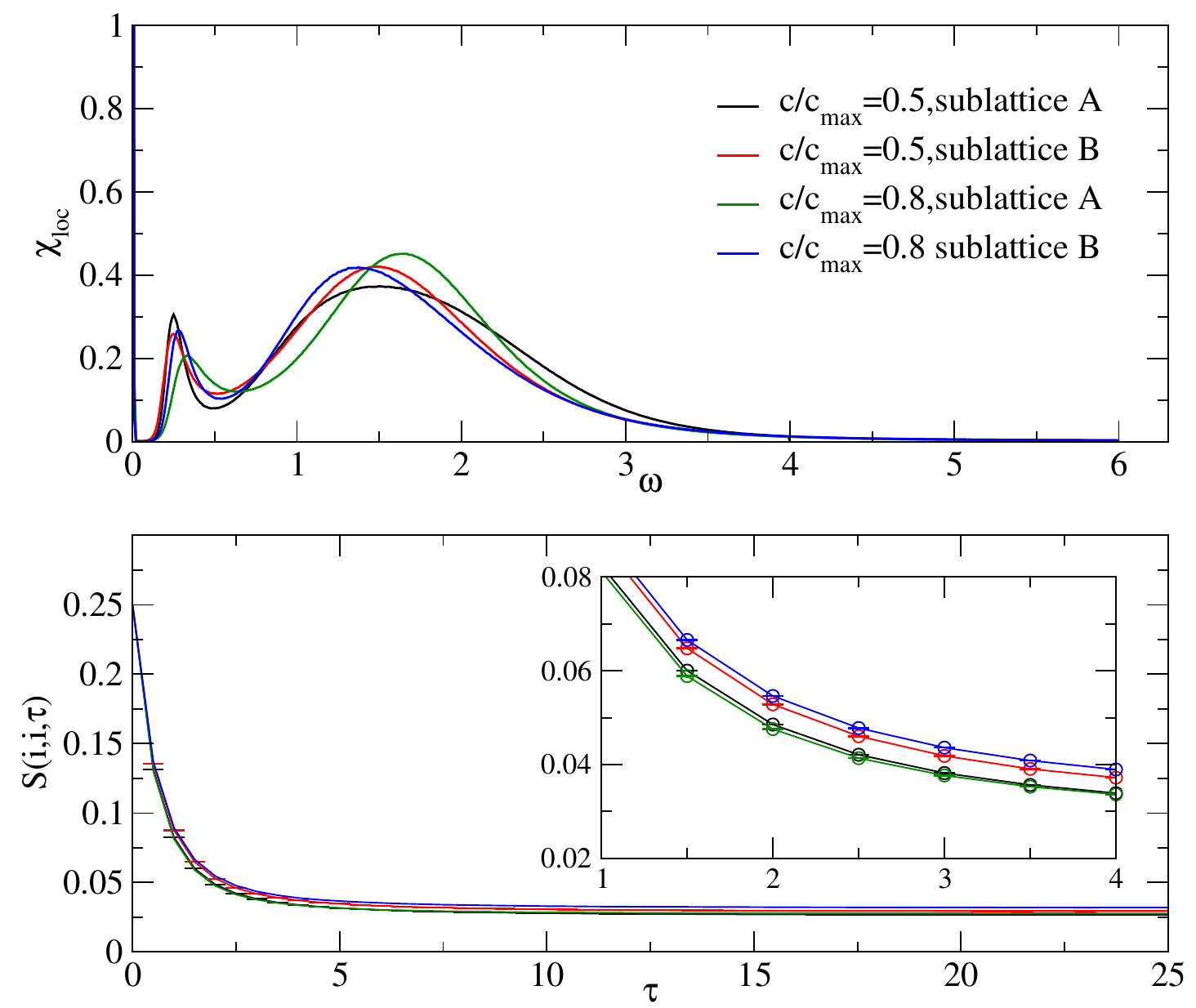} \caption{(a) The real-frequency local susceptibility obtained by numerical analytical continuation. (b) The imaginary time correlation of the spins defined in Eq.(25) as a function of $\tau$. Since $S(i,i,\tau)$ is symmetric to $\beta/2$, only the curves in the range $[0,\frac{\beta}{2}]$ are plotted. Here the inverse temperature is $\beta=50$.}
\label{fig8}
\end{figure}

\section{The strained $XY$ antiferromagnet}

\begin{figure}[htbp]
\centering \includegraphics[width=8.8cm]{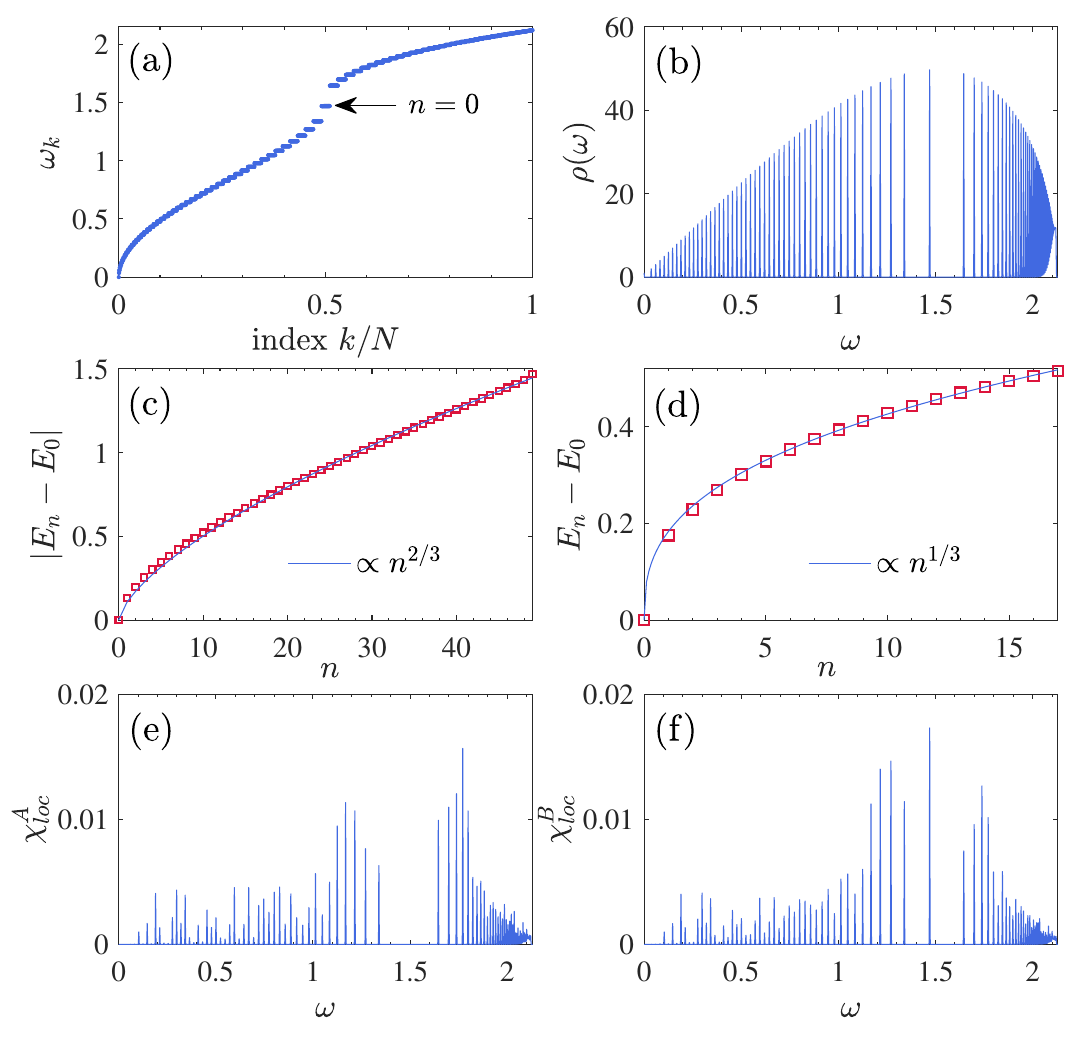} \caption{ (a) The magnon spectrum of the strained $XY$ Hamiltonian. (b) The magnon density of states for the spectrum in (a). The quantized energies measured from the $0$th PLLs: (c) those below the $0$th PLLs; (d) those above the $0$th PLLs. The data in (c) and (d) are best fitted by $|E_n-E_0|\propto n^{\frac{2}{3}}$ and $n^{\frac{1}{3}}$, respectively. The spin-wave local susceptibilities $S(i,i,\omega)$ on: (e) sublattice A; (f) sublattice B. In all figures, the strain strength is $c/c_{max}=1$.}
\label{fig9}
\end{figure}

We next consider the spin-$\frac{1}{2}$ $XY$ antiferromagnetic Hamiltonian described by
\begin{align}
H_{XY}=J\sum_{\langle ij\rangle}(S_i^xS_j^x+S_i^yS_j^y).
\end{align}
The above model becomes the $XZ$ Hamiltonian by a $90$ degree clockwise rotation of the $XYZ$ coordinate system about the $x$-axis. The resulting Hamiltonian writes as\cite{Joannopoulos1987}
\begin{align}
H_{XZ}=&J\sum_{\langle ij\rangle}(S_i^xS_j^x+S_i^zS_j^z),  \\ \nonumber
=&J\sum_{\langle ij\rangle}(S_i^zS_j^z+\frac{1}{4}\sum_{\mu,\nu=\pm}S_i^{\mu}S_j^{\nu}).
\end{align}
Under a triaxial strain, the same modulation of the exchange coupling with that in Eq.(2) can be made. Then the application of LSWT to $H_{XZ}$ is straightforward, and the procedure is similar to that described in Sec.III.

Figure 9(a) plots the magnon spectrum of the $XZ$ Hamiltonian Eq.(28). In the presence of strain, the energy levels also become discrete and degenerate sets, resulting sharp peaks in the magnon density of states [see Fig.9(b)]. Unlike the Heisenberg case, the $0$th PLL now appears in the middle of the spectrum, and the peaks are no longer equally-spaced. We take the energy of the $0$th PLL as the reference, and plot the $n$th-PLL energy as a function of $n$ in Fig.9(c) and (d). We find the curve is best fitted by $|E_n-E_0|\propto n^{\frac{1}{3}} (n^{\frac{2}{3}})$ for PLLs above (below) the $0$th one. This relation is quite different from the ones in the Heisenberg model and the graphene, representing a new kind of Landau quantization. We also demonstrate the local susceptibility at the sample center. As shown in Fig.9(e) and (f), $\chi_{loc}^{A}$ and $\chi_{loc}^{B}$ exhibit the same oscillating behavior with the magnon density of states. However it is noted that now the $0$th PLL is totally located on sublattice B, which is in sharp contrast to the Heisenberg case where it is on sublattice A.

\begin{figure}[htbp]
\centering \includegraphics[width=7.5cm]{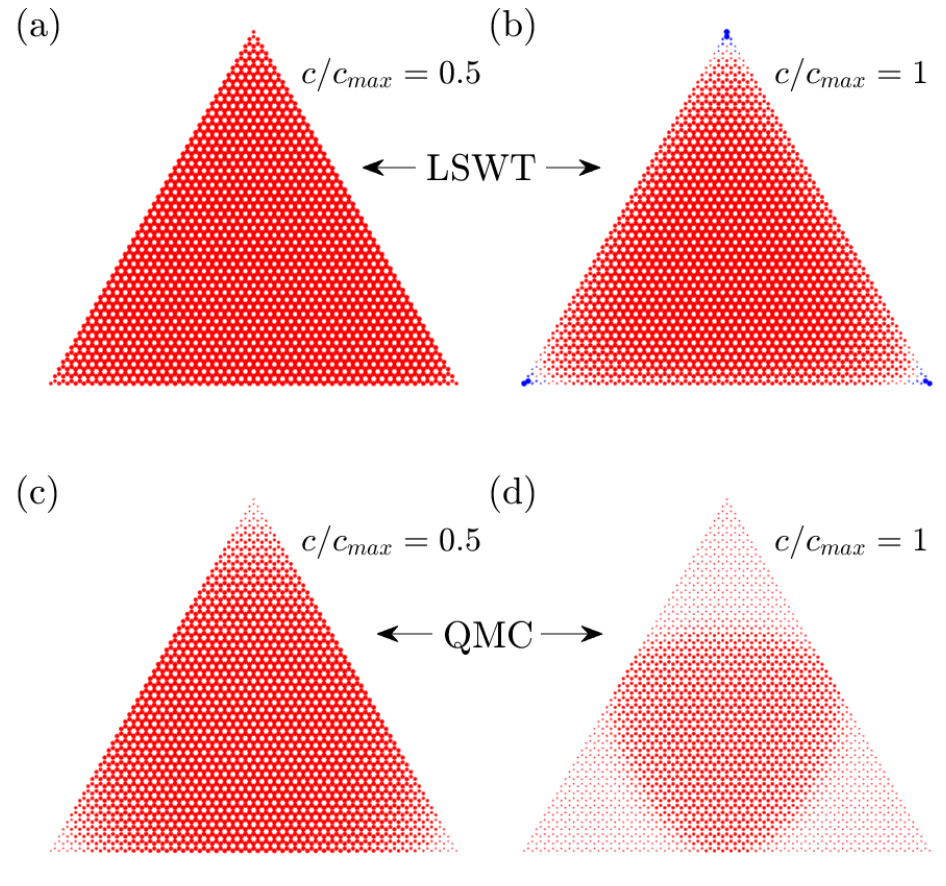} \caption{ The distribution of the local magnetization obtained
with LSWT at (a) $c/c_{max}=0.5$ and (b) $c/c_{max}=1$. The distribution of $\tilde{m}_s(i)$ calculated with QMC at the strain strengths (c) $0.5$ and (d) $1$. }
\label{fig10}
\end{figure}

To show the effect of the strain on the AF order, we calculate the distribution of the local magnetization within the LSWT at the strain strengthes $c/c_{max}=0.5,1$. As shown in Fig.10(a) and (b), the AF order persists in the central region, and vanishes in the regions near the corners after the strain is applied. Besides, the size of the central AF region decreases with increasing the strain. Although the above results are similar to the Heisenberg case, the persisted AF region is much larger than that in the strained Heisenberg model for the same strain strength. Both the Heisenberg and $XZ$ models have exactly the same dimerization of the couplings induced by the strain. Hence the result implies the AF order in the $XZ$ model is more robust to the dimerization, which is also found in our recent work\cite{guo2020quantum}. Qualitatively, the Heisenberg Hamiltonian contains three spin components compared to two ones in the $XY$ model, thus the quantum fluctuations are much stronger in the Heisenberg case.  The competition between the AF order and the dimerization is mainly determined by the spin quantum fluctuations. Due to the strong quantum fluctuations, the AF order in the Heisenberg model is broken by relatively weak dimerization. The dimerizaiton is gradually enhanced when approaching the corners. Hence the AF breakdown in the Heisenberg case happens deeper inside the lattice, resulting in a smaller AF region.

The related quantity defined in Eq.(20) is also calculated using the QMC methods at $c/c_{max}=0.5,1$, which is shown in Fig.10(c) and (d). While the QMC results are qualitatively consistent with those from LSWT, the exact QMC simulations predict much smaller AF regions for the same strain strengths.

\section{Conclusions}

The evolution of AF order and the verification of the magnon PLLs of the strained honeycomb antiferromagnets are studied using the QMC simulations on a triangular geometry. After the strain is applied, the AF order persists in the central region of the geometry, and is gradually broken starting from the corners. While LSWT gives the correct trend of the AF evolution, it overestimates the AF region than the exact QMC result for the same strain strength. The $XY$ Hamiltonian has similar AF evolution with strain, except that the persisted AF region is much
larger than that of the corresponding Heisenberg model. Besides, the properties of the spin-wave magnon PLLs are quite different in the $XY$ model: the $0$th Landau level appears in the middle of the spectrum, and the quantized energies above (below) it are proportional to $n^{\frac{1}{3}} (n^{\frac{2}{3}})$. For the Heisenberg case, we extract the real-frequency local susceptibility from the imaginary time correlations of the spins using SAC, and find no sign of the magnon PLLs. Due to the intrinsic problem of numerical analytical continuation, the existence or not of PLLs is still unsure, thus represents an interesting open question.

Recently significant progress has been made in two-dimensional (2D) quantum magnetic materials\cite{burch2018magnetism,gibertini2019magnetic,PhysRevX.10.011062}. The easily cleavable layered magnetic van der Waals materials allow the creation of atomically thin layers. Among the many discovered materials, monolayer of metal phosphorus trichalcogenides ($\textrm{MPX}_3$) show long-range AF order with a honeycomb geometry\cite{PhysRevB.94.184428,mak2019probing}. Hence the triaxial strain may be engineered based on these 2D honeycomb antiferromagnets, and it is very possible that our results are studied experimentally.

\section*{Acknowledgments}
The authors thank Shoushu Gong, Tianyu Liu, Hui Shao, Yancheng Wang, Wen Yang, Chenyue Wen, Xingchuan Zhu for helpful discussions. J.S and H.G. acknowledges support from the NSFC grant Nos.~11774019 and 12074022, the NSAF grant in NSFC with grant No. U1930402, the Fundamental Research
Funds for the Central Universities and the HPC resources
at Beihang University. N. M. is supported by the NSFC grant No.12004020.
S.F. is supported by the National Key Research and Development Program of China under Grant No. 2016YFA0300304, and NSFC under Grant Nos. 11974051 and 11734002.

\appendix

\renewcommand{\thefigure}{A\arabic{figure}}

\setcounter{figure}{0}

\section{The details of some derivations in LSWT}
We first outline the method to obtain the Bogliubov transformation $T$ briefly. Due to the commutation relations of the bosonic operators, we have
\begin{align}
T\Gamma T^{\dagger}=\Gamma,
\end{align}
where the matrix $\Gamma$ is given by
\begin{align}
\Gamma=\left(
         \begin{array}{cc}
           I_{N_A} & 0 \\
           0 & -I_{N_B} \\
         \end{array}
       \right),
\end{align}
with $I_{N}$ the $N\times N$ identity matrix, and satisfies $\Gamma^2=1$.
Since $M$ is diagonalized by $T$, i.e., $T^{\dagger}MT=D$ [the diagonal matrix $D=\rm{diag}(\omega_1,...,\omega_{N_s})$], the following relation is reached with Eq.(A1),
\begin{align}
\Gamma MT=T\Gamma D.
\end{align}
The eigenvectors of $\Gamma M$ are the transformation $T$, which is directly obtained by diagonalizing $\Gamma M$. Suppose the diagonal matrix of the eigenvalues of $\Gamma M$ is $D'$. The magnon spectrum is given by the diagonal elements of the matrix $D=\Gamma D'$.

Generally, the eigenvector matrix $T$ from a direct diagonalization may not be normalized. We can calculate $L=T^{\dagger}\Gamma T$, and get a diagonal matrix $L={\rm{diag}}(l_1,...,l_{N_s})$. Then the normalized matrix is constructed as
\begin{align}
L_2={\rm{diag}}(\sqrt{|l_1|},...,\sqrt{|l_{N_s}|}).
\end{align}
The normalized transformation is obtained subsequently by the matrix product $TL_2$.

We next show the details in deducing the expression for the local spin susceptibility.
For the site on sublattice A and within the LSWT, $\chi^{A}_{loc}$ is given by,
\begin{align}
\chi^{A}_{loc}&=\int_{-\infty}^{\infty} d t e^{i \omega t} \frac{1}{2}\left\langle S_{i}^{+}(t) S_{i}^{-}(0)+S_{i}^{-}(t) S_{i}^{+}(0)\right\rangle \\ \nonumber
&=S \int_{-\infty}^{\infty} d t e^{i \omega t}\left\langle a_{i}(t) a_{i}^{\dagger}+a_{i}^{\dagger}(t) a_{i}\right\rangle.
\end{align}
Using the Bogliubov transformation, we have
\begin{align}
&a_{i}(t) a_{i}^{\dagger}+a_{i}^{\dagger}(t) a_{i} \\ \nonumber
=&\left(\sum_{k=1}^{N_A}A_{+,i k} \alpha_{k}(t)+\sum_{i=1}^{N_B}A_{-,i k} \beta_{k}^{\dagger}(t)\right)\\ \nonumber
&\left(\sum_{k=1}^{N_A}A_{+,j k}^{*} \alpha_{k}^{\dagger}+\sum_{i=1}^{N_B}A_{-,j k}^{*} \beta_{k}\right) \\ \nonumber
+&\left(\sum_{k=1}^{N_A}A_{+,j k}^{*} \alpha_{k}^{\dagger}(t)+\sum_{i=1}^{N_B}A_{-,j k}^{*} \beta_{k}(t)\right)\\ \nonumber
&\left(\sum_{k=1}^{N_A}A_{+,i k} \alpha_{k}+\sum_{i=1}^{N_B}A_{-,i k} \beta_{k}^{\dagger}\right) \\ \nonumber
=&\sum_{i=1}^{N_A}A_{+,i k}A_{+,i k}^{*}\langle \alpha_{k}(t)\alpha_{k}^{\dagger}\rangle  \\ \nonumber
+&\sum_{i=1}^{N_B}A_{-,i k}A_{-,i k}^{*}\langle \beta_{k}(t)\beta_{k}^{\dagger}\rangle,
\end{align}
where the last step is obtained by considering the zero temperature and all other averages of quadratic operators are zero.
With the definition of time-dependent operators $C(t)=e^{iHt}Ce^{-iHt}$ ($C=\alpha,\beta$), we obatain
\begin{align}
\langle C(t)C^{\dagger}(0)\rangle=e^{-i\omega t},
\end{align}
where $\omega$ is the eigenenergy of $C$-type quasiparticle.
Substituting Eq.(A7) into Eq.(A6), we reach the expression for the local susceptibility,
\begin{align}
\chi^{A}_{loc}=\int_{-\infty}^{\infty} d t e^{i \omega t} (&\sum_{k=1}^{N_B}A_{-,i k} A_{-,i k}^{*}e^{-i \omega_{k} t}\\ \nonumber
+&\sum_{k=1}^{N_A}A_{+,i k}^{*} A_{+,i k} e^{-i \omega_{k} t} ),
\end{align}
\begin{align}
\chi^{A}_{loc}=&\sum_{k=1}^{N_B}A_{-,ik} A_{-,i k}^{*}\delta\left(\omega-\omega_{k}\right)\\ \nonumber
              +&\sum_{k=1}^{N_A}A_{+,i k}^{*} A_{+,i k} \delta\left(\omega-\omega_{k}\right).
\end{align}
The expression for $\chi^{B}_{loc}$ can be deduced in a similar way.

\section{More LSWT and QMC results}

\begin{figure}[htbp]
\centering \includegraphics[width=7.5cm]{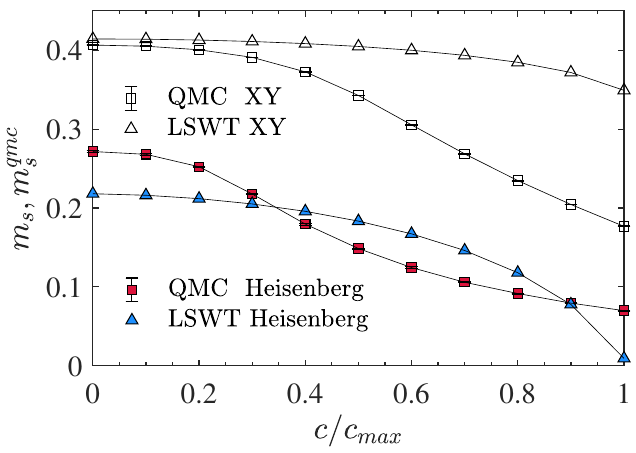} \caption{The mean magnetizations $m_s$(LSWT) and $m_s^{qmc}$(QMC) as a function of strain strength for the $XY$ and Heisenberg models. }
\label{afig1}
\end{figure}

Figure \ref{afig1} plots the mean magnetizations as a function of strain strength for the $XY$ and Heisenberg models. In both figures, the LSWT and QMC results are qualitatively consistent, and the curves all decrease monotonically with the strain. At the same strain strength, the value the $XY$ case is much larger than that of the Heisenberg model, which implies the antiferromagnetism is more robust to the strain in the $XY$ model.

\begin{figure}[htbp]
\centering \includegraphics[width=7.5cm]{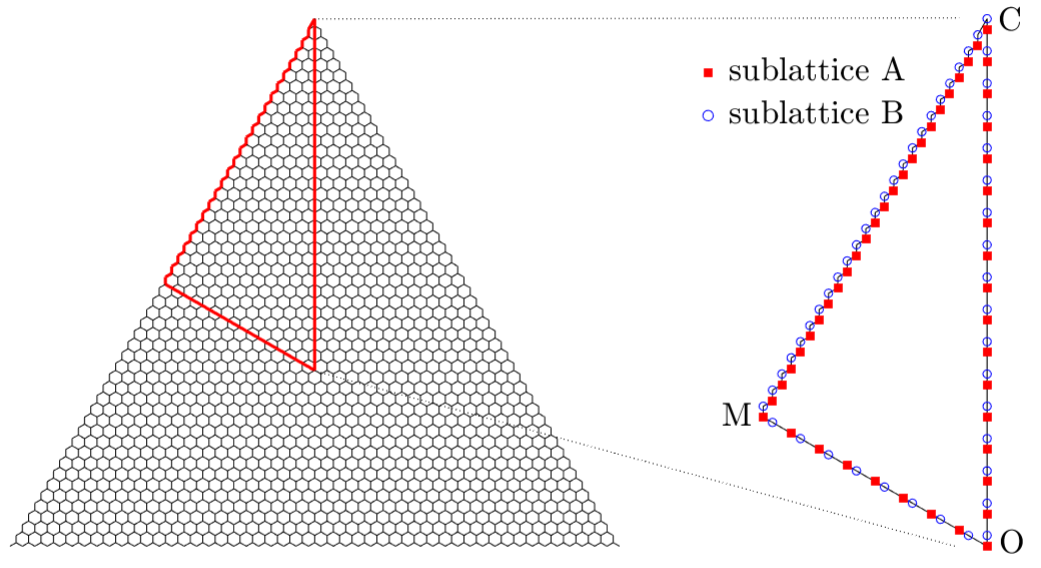} \caption{Schematical illustration of the high-symmetry paths, along which the values of the local magnetizations will be demonstrated in the following two figures. }
\label{afig2}
\end{figure}

\begin{figure}[htbp]
\centering \includegraphics[width=8.5cm]{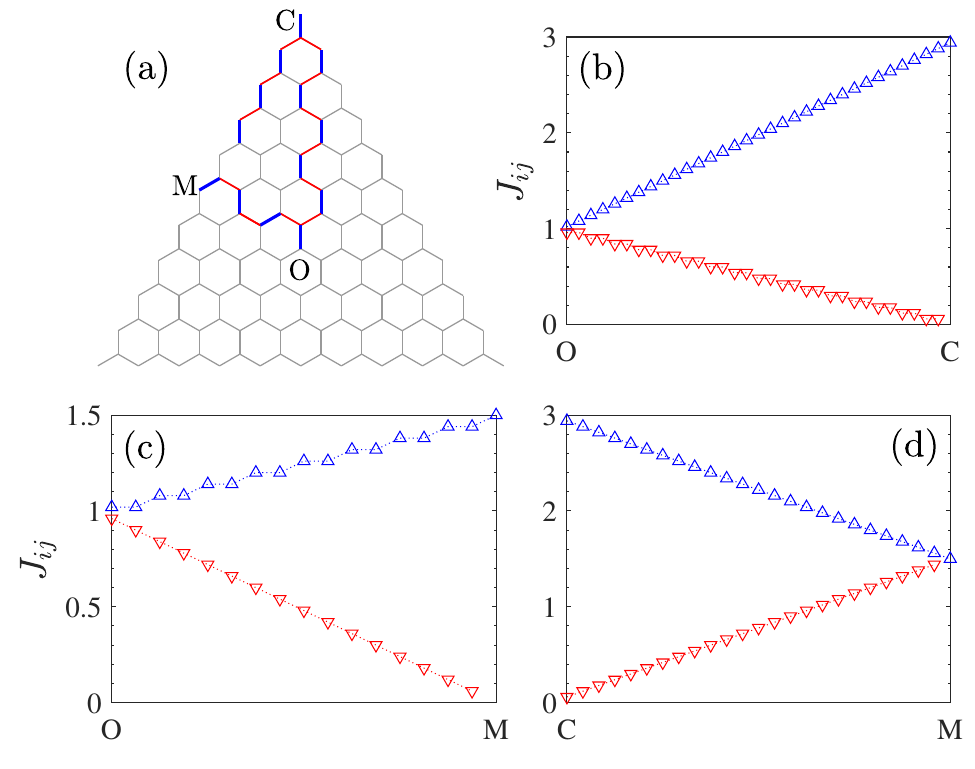} \caption{(a) Schematical illustration of the bonds along the high-symmetry paths. The enhanced (weakened) bonds on each path are marked by the thick blue (thin red) solid line. The values of the exchange couplings $J_{ij}$ along the path: (b) $\textrm{O}\rightarrow \textrm{C}$; (c) $\textrm{O}\rightarrow \textrm{M}$; (d) $\textrm{C}\rightarrow \textrm{M}$}
\label{afig3}
\end{figure}

Figure \ref{afig3} plots the values of the exchange couplings along the high-symmetry paths. The dimerizations are formed on these one-dimensional chains. Generally, the modified value is proportional to the distance from the lattice center. Thus the bonds are continuously enhanced or weakened when being away from the center, generating the strongest dimerization near the corner. Among all the modified bonds, the ones vertically connecting the zigzag edges have the smallest values. In contrast, the exchange couplings are largest on the bonds both near the corners and with the orientation vertical to the opposite edge. It is also noted that the bonds is less changed in the central region, where the pseudo-magnetic field is the most uniform.

\begin{figure}[htbp]
\centering \includegraphics[width=7.5cm]{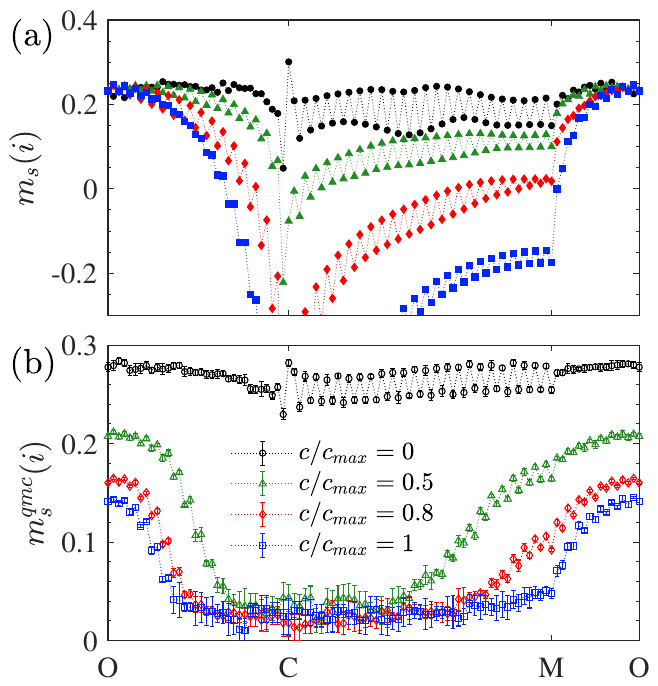} \caption{The local magnetizations along high-symmetry lines at various strain strengths in the Heisenberg model: (a) the LSWT results; (b) the QMC results. }
\label{afig4}
\end{figure}

\begin{figure}[htbp]
\centering \includegraphics[width=7.5cm]{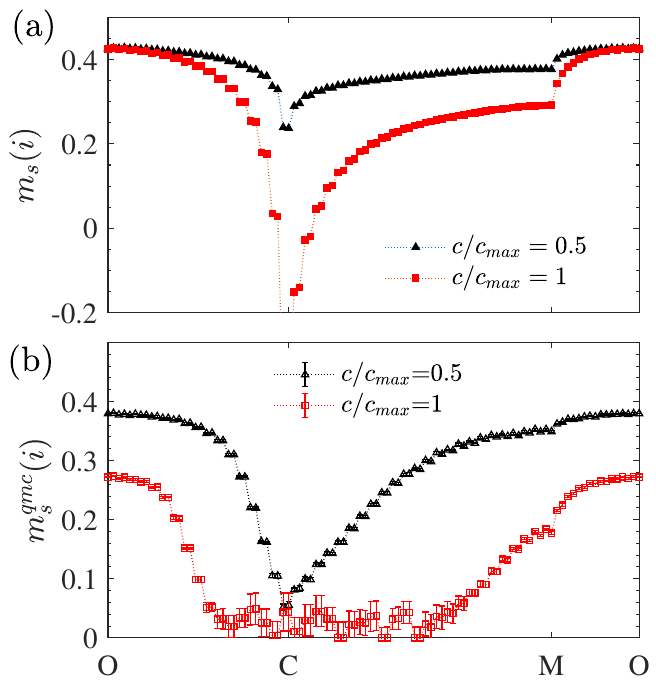} \caption{The local magnetizations along high-symmetry lines at various strain strengths in the $XY$ model: (a) the LSWT results; (b) the QMC results. }
\label{afig5}
\end{figure}

\begin{figure}[htbp]
\centering \includegraphics[width=7.5cm]{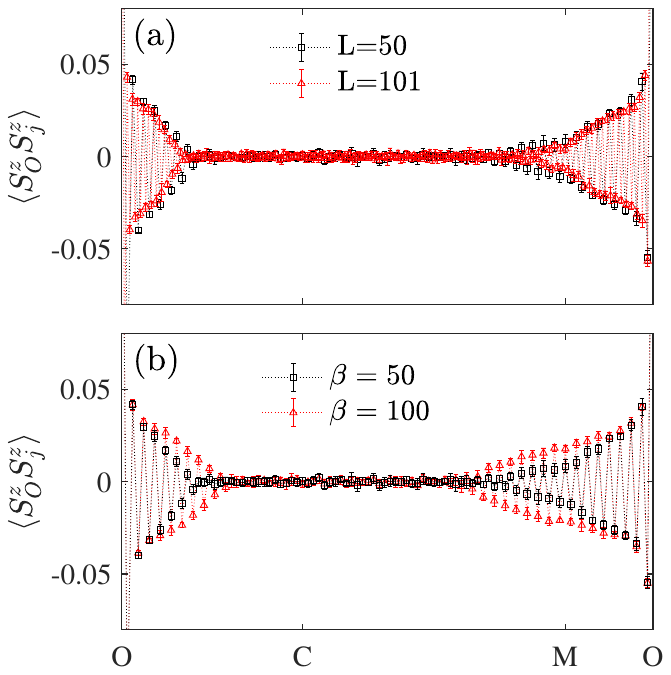} \caption{The QMC correlations between the spin on the central point $O$ and the spins on the sites along the high-symmetry paths in the strained Heisenberg model: (a) the linear size $L=50,101$ at the inverse temperature $\beta=50$; (b) $\beta=50,100$ with $L=50$. The strain strength is $c/c_{max}=0.8$. }
\label{afig6}
\end{figure}

\begin{figure}[htbp]
\centering \includegraphics[width=7.5cm]{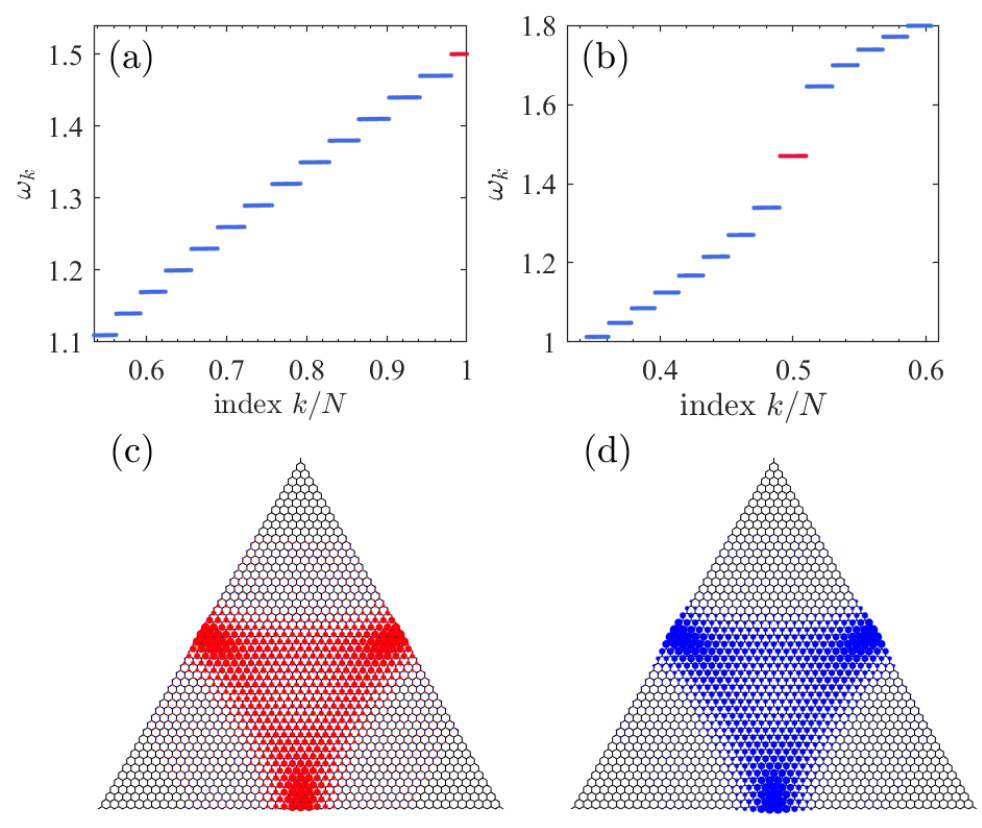} \caption{The spin-wave magnon spectrums and the distributions of the wave functions of the $0$th PLLs in: (a) and (c) the Heisenberg model; (b) and (d) the $XY$ model. In (c) and (d), the value on each site is represented by the radii of a solid circle, whose color distinguishes the sublattice. The red (blue) color corresponds to sublattice A (B). The wave functions of the $0$th PLLs distribute only on one kinds of sublattice, which is opposite for the two considered Hamiltonians. The strain strength is $c/c_{max}=1$.}
\label{afig7}
\end{figure}

Corresponding to Fig.2 and Fig.5 in the main text, Figure \ref{afig4} plot the values of the local magnetizations along the high-symmetry lines of the triangular geometry. In the absence of strain, both the LSWT and QMC show the sites on the zigzag open boundary of the geometry are most affected. The values on the two sublattices have clear difference. After the strain is applied, this difference is greatly suppressed in the QMC results, but still evident in the LSWT ones. The local magnetization takes the maximum value at the central point $O$ (see Fig.\ref{afig4}), and continuously decreases as the sites moves to the corner point $C$ and the middle point $M$ of the boundary. This is consistent the evolution of the AF order discussed in the main text. For all considered strain strengths, the values along $O-M$ keep finite, thus the AF order therein preserves. However for the other two paths, there are clear transition points, whose positions vary with the strain strength, resulting larger nonmagnetic regions for stronger strains.
While the LSWT values change little near the point $O$, the QMC ones gradually decrease with the strain. The values at the point $C$ become negative and extremely large for strong strains, which is unrealistic and suggests the LSWT fails here. Since the magnon PLLs only become apparent at strong strains, their existence predicted by the LSWT is highly questionable.

We plot the local magnetizations in the same way for the strained $XY$ Hamiltonian. As shown in Fig.\ref{afig5}, the results are similar except the values are much larger, and the AF order persists in a larger region for the same strain strength.

We also calculate the spin correlations on larger lattice size and lower temperature, as shown in Fig. \ref{afig6}. When the size is doubled, the values decrease slightly. We then compare the results at $\beta=L,2L (L=50)$. The nonzero values on the paths become larger at lower temperature, and the magnetic transition points move to the corner point $C$. These trends imply the central AF region will persist in the thermaldynamic limit and zero temperature.

In the strained graphene, the $0$th PLL only resides on one kind of sublattice\cite{PhysRevB.90.155418}, which is determined by that of the outmost sites of the zigzag boundaries (here it is sublattice B). Figure \ref{afig7} shows the distributions of the wave functions of the $0$th PLLs. Although the $0$th magnon PLL of the $XY$ and Heisenberg cases also resides only on one kind of sublattice, it is sublattice A (B) for the Heisenberg ($XY$) Hamiltonian. In this sense, the magnons in the $XY$ model behave more like Dirac fermions in graphene.

\begin{figure}[htbp]
\centering \includegraphics[width=8.cm]{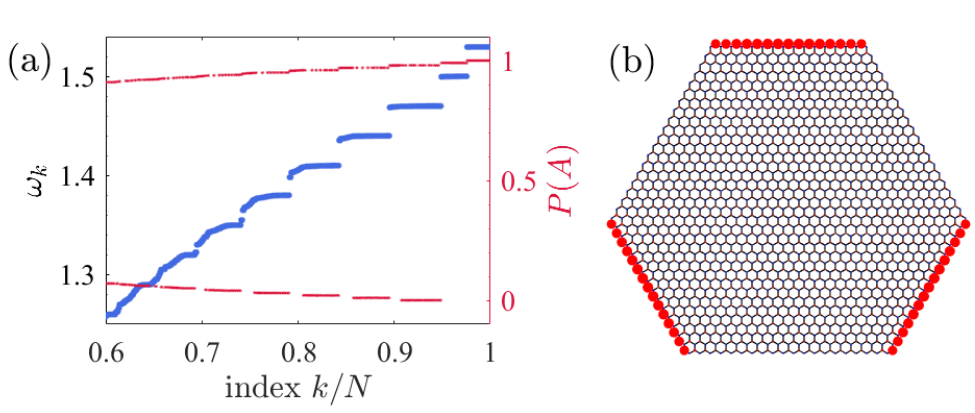} \caption{(a) The magnon spectrum near the high-energy end and the weight of the corresponding eigenstates on the $A$ sublattice (right axis) in the Heisenberg model. (b) The total probability density of the degenerate states associated to the highest eigenenergy in (a). The strain strength is $c/c_{max}=1$.}
\label{afig8}
\end{figure}

The feature of the strained honeycomb antiferromagnet is dependent on the geometry. Based on a $N=50$ triangular geometry, we cut off three $N_1=15$ small triangles from the corners, and generate a hexagonal flake with zigzag edges, among which the three new edges are terminated by
$N_1$ $A$ sites and the other ones are terminated by $N-2N_1$ $B$ sites. Figure \ref{afig8}(a) shows the magnon spectrum of the strained Heisenberg model in this hexagonal geometry. While the original PLLs are almost unchanged, there appear additional $3N_1$ degenerate states right above the $0$th PLL, whose probability densities are totally distributed on $A$ sites near the three new edges, exhibiting
some kind of topological signature. Although this property is very similar to that found in the
strained graphene\cite{PhysRevB.90.155418,PhysRevX.4.021042}, clear differences exist, such as: these edge states are sublattice polarized and they are not degenerate with the $0$th PLL. We leave a detailed study of this issue for future work.

\bibliography{ddirac}

\end{document}